\documentclass{tMOP2e}

\usepackage{color}

\usepackage{subfigure}
\usepackage{braket}

\begin{document}
\title{{Bichromatic State-insensitive Trapping of Caesium Atoms} }

\author{M.M. Metbulut$^{\ast}$\thanks{$^\ast$Corresponding author. Email: ucapmmm@ucl.ac.uk 
\vspace{6pt}} and F. Renzoni\\\vspace{6pt}  {\em{Department of Physics and Astronomy, University College London, London, UK}}\\
\received{v4.1 released September 2014} }
\maketitle

\begin{abstract}
State-insensitive dipole trapping of multilevel atoms can be achieved by an appropriate choice of the wavelength of the trapping laser, so that the interaction with the different transitions results in equal AC Stark shifts for the ground and excited states of interest. However this approach is severely limited by the availability of  coherent sources at the required wavelength and of appropriate power. This work investigates state-insensitive trapping of caesium atoms for which the required wavelength of 935.6 nm  is inconvenient in terms of experimental realization. Bichromatic state-insensitive trapping is proposed to overcome the lack of suitable laser sources. We first consider pairs of laser wavelengths in the ratio 1:2 and 1:3, as obtained via second- and third- harmonic generation. We found that the wavelength combinations $931.8-1863.6$ nm and $927.5-2782.5$ nm are suitable for state-insensitive trapping of caesium atoms.  In addition, we examine bichromatic state-insensitive trapping produced by pairs of laser wavelengths corresponding to  currently available high power lasers. These wavelength pairs were found to be  in the range of $585-588$ nm and $623-629$ for one laser and $1064-1080$ nm  for the other.


\end{abstract}

\section{Introduction}

Dipole trapping is of great importance for cold atom research. The ability to manipulate trapped neutral atoms plays an important role in many experiments in precision spectroscopy \cite{ye2008}, optical clocks \cite{kat09,tak05,lud08}, quantum control \cite{yem} and quantum information processing \cite{qin}. Dipole trapping is produced by a spatial intensity gradient of the trapping laser field(s), which leads to a trapping potential for the atomic ground state. In all the experiments involving the interrogation of an optical transition, as for example in precision spectroscopy, the effect of the trapping light on the transition has to be taken into account. In general, the AC Stark shifts of the ground and excited states of the optical transition of interest are different, thus the trapping light leads to the formation of a differential shift between the two states. Such a differential AC Stark shift, which is position-dependent via the position-dependence of the laser intensity, has a number of drawbacks. It causes decoherence, thus limiting the applicability of dipole trapping in experiments requiring long coherence times \cite{decoh}. It makes it difficult to perform precision spectroscopy on trapped atoms, as it results in a line shift and its position-dependence causes a line broadening. This also complicates the continuous monitoring of trapped atoms. Moreover, free-space cooling mechanisms cannot operate on trapped atoms due to the differential AC Stark shift. This also introduces difficulties in efficient loading of atoms into dipole traps. 

In 1999, in the context of optical lattice clocks, Katori \emph{et} al. \cite{katori1999optimal} introduced a technique to keep the internal state of a trapped atom unaffected by the trapping light.  By taking advantage of the multi-level structure of atoms, it was proposed to tune the trapping laser to a specific wavelength - the so called {\it magic} wavelength - at which the differential light shift between the excited and the ground states of a particular transition vanishes. This is the result of the interaction of the dipole trap laser field with all the different transitions of the multilevel atom, with relative weights determined by the laser detuning. The effect was demonstrated for a Strontium optical lattice clock, with state-insensitive trapping allowing for simultaneous laser cooling and dipole trapping.  Later in $2003$, state-insensitive trapping of caesium atoms confined in an optical cavity field at the magic wavelength of $935.6$ nm was demonstrated by McKeever \emph{et} al. \cite{mckeeverpaper} in the context of cavity quantum electrodynamics. It was shown that continuous monitoring of trapped atoms becomes possible with state-insensitive trapping. Later in $2010$, state-insensitive trapping of caesium atoms in free space was also demonstrated \cite{boeing}. However, in this case the size and depth of the trap is severely limited by the power of available laser sources at the required wavelength.


The magic wavelengths for several other neutral atoms, such as Mg \cite{mg07}, Hg \cite{hg11}, Rb \cite{rb10}, Ba \cite{ba12}, Ra \cite{ba12}, Ca \cite{ca04} and Li \cite{li} have been determined. Magic wavelengths for Ryberg atoms have also been calculated \cite{rydberg1,rydberg2} However in terms of experimental realizations of  state-insensitive trapping, some of the identified magic wavelengths are of limited practical use. More specifically, the calculated magic wavelengths may be in an inappropriate frequency range for current laser technology,  either in terms of wavelength or  power required. In the present work bichromatic trapping is investigated to overcome the abovelisted problems associated with monochromatic state-insensitive trapping. Bichromatic trapping  was first proposed by Arora {\it et al.} for rubidium atoms \cite{arora},  as the magic wavelength for monochromatic trapping was inconvenient due to the atomic polarizability being too small for efficient trapping. In contrast to rubidium, the major obstacle for state-insensitive monochromatic trapping of caesium atoms in free space is the lack of laser sources with sufficiently high power. Motivated by this, we investigate bichromatic state-insensitive trapping of caesium atoms to determine combinations of magic wavelengths more convenient than the monochromatic magic wavelength of $935.6$ nm.

Bichromatic state-insensitive trapping relies on the combined action of two independent laser fields with different wavelengths. One of the lasers is used as the trap laser and the other one is used as a control laser to compensate for the differential light shift between the ground and excited states of the atomic  transition of interest. Bichromatic state-insensitive trapping is more flexible that its monochromatic counterpart, as several different combinations of parameters of the two lasers may lead to state-independent trapping. Thus, bichromatic trapping may be useful to extend the range of possible wavelengths that can be used for state-insensitive trapping.

This paper is organised as follows. First, the case of  laser wavelengths  in the 1:2 and 1:3 ratio is considered, with particular attention to the infrared region of spectrum. This corresponds to pairs of laser fields with one obtained from the other one via second- or third-harmonic generation. Next, wavelength combinations attainable by currently available high power lasers were considered, and appropriate combinations for state-insensitive trapping determined. Finally, the effect of unwanted  laser intensity and wavelength variations is discussed.

\section{Numerical Results}

We investigate bichromatic schemes for caesium atoms, which provide state-insensitive trapping with respect to the D$_2$ line transition, i.e. with no differential light shift for that transition.

Our numerical approach follows standard procedure.  We evaluate separately the AC Stark shift of the ground state $6S_{1/2}(F=4)$ and the excited state $6P_{3/2}(F=5)$ in the presence of both trap and control lasers as functions of the laser fields' wavelengths. This is done  by using time-independent perturbation theory. Then we determine the crossing point of the AC Stark shifts for the two states. This gives the wavelength combination required for state-independent trapping. All the atomic hyperfine levels and their corresponding Zeeman sublevels are  included in our calculation. In this way,  we can identify the required  combination of magic wavelengths  for situations in which the trapped atoms are prepared in specific Zeeman sublevels. 

Consider an atom driven by a laser field with intensity $I_L$, and frequency $\omega_{0}$. 
The AC Stark shift experienced by the atomic level $\ket{i} = \ket{F_i m_i}$ as a result of the laser coupling  to a set of states $\ket{j} = \ket{F_j m_j}$ is given by

\begin{equation}
\Delta E_i =(2F_{i}+1) I_L\sum_{j}{\frac{3\pi c^{2} A_{ij}  }{2\Delta\omega_{0}^{3}   }(2F_{j}+1)(2J_{j}+1)\begin{pmatrix} F_{j} & 1 & F_{i} \\ m_{j} & q & -m_{i} \end{pmatrix}^{2}\begin{Bmatrix} J_{i} & J_{j} & 1 \\ F_{j} & F_{i} & I \end{Bmatrix}^{2}},
\label{actot}
\end{equation} 
where  $ A_{ij}$ is the Einstein A-coefficient for the transition $\ket{i}\leftrightarrow\ket{j}$, $I$ the nuclear spin and $\Delta$ the laser field detuning from resonance with the considered transition. The parameter $q$ parametrizes to the polarization of the light: $q=\pm 1$ for $\sigma^{\pm}$ light and $q=0$ for linearly polarized light.

The AC Stark shift produced by the trap laser ($\Delta E_i^t$) and control laser  ($\Delta E_i^c$) are calculated separately via Equation  \ref{actot}, with the total AC Stark shift given by their sum.

All calculations presented in this work are carried out for linear polarization ($q=0$) of the trap and control laser fields. This is the most widely used configuration for dipole trapping, as the ground state Zeeman sublevels remain degenerate in the presence of the trapping light.  Calculations of the AC Stark shift of the $\ket{i}$ states  $6S_{1/2}(F=4)$ and $6P_{3/2}(F=5)$ are performed, including contributions from the coupling with the $\ket{j}$ states  up to $15$S, $11$P and $11$D. 

\subsection{Magic Wavelength Pairs for $\lambda_{\rm c}=2\lambda_{\rm t}$ and $\lambda_{\rm c}=3\lambda_{\rm t}$}

\begin{figure}[h]
\begin{center}
\begin{minipage}{140mm}
\hspace*{-1.3cm}\subfigure[$\lambda_{control}=2\lambda_{trap}$.]{
\resizebox*{8.5cm}{!}{\includegraphics{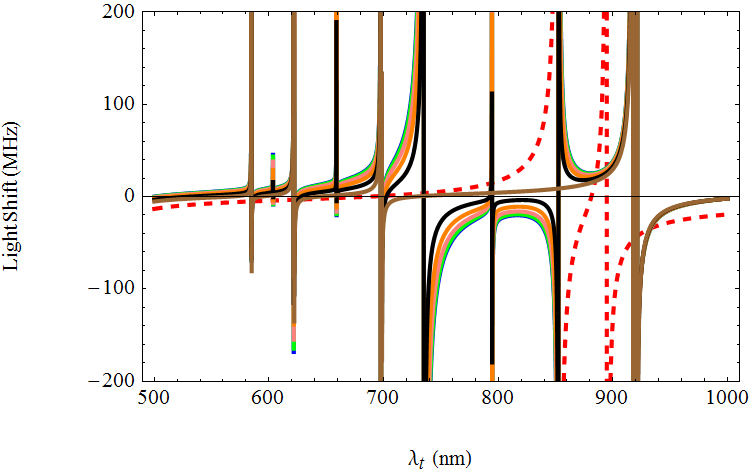}}}\hspace{6pt}
\subfigure[$\lambda_{control}=3\lambda_{trap}$.]{
\resizebox*{8.5cm}{!}{\includegraphics{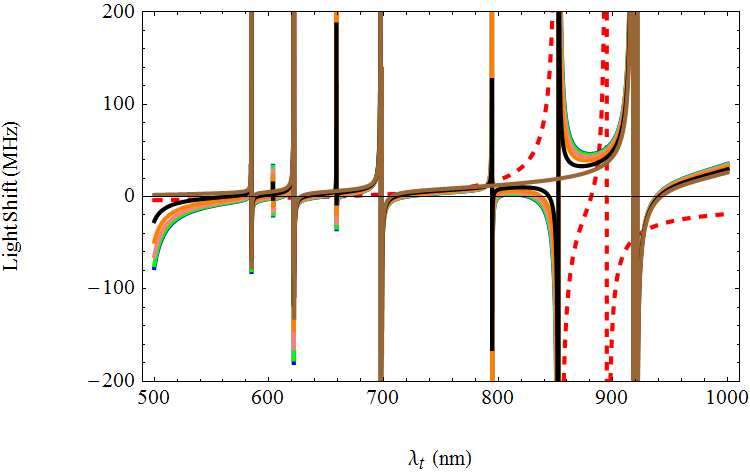}}}
\caption{AC Stark shift for the $F_{g}=4\rightarrow F_{e}=5$ transition as function of trap laser wavelength for (a) $\lambda_{\rm c}=2\lambda_{\rm t}$ and (b) $\lambda_{\rm c}=3\lambda_{\rm t}$. The intensities of the trap and control laser is equal and kept at $2\times{10}^{9}$ W/$m^{2}$. 
Red dashed line is the ground state $6S_{1/2}$, blue, green, pink, orange, black and brown lines are the excited state $6P_{3/2}$ with $m=0$, $m=\pm 1$, $m=\pm 2$, $m=\pm 3$, $m=\pm 4$ and $m=\pm 5$ Zeeman sublevels respectively.}
\label{fig lcont=2ltrapeqint1}
\end{minipage}
\end{center}
\end{figure}

We consider here bichromatic trapping schemes with the trap and control laser wavelengths in the ratio 2:1 and 3:1, as obtained from a single laser source via second-  or third-harmonic generation, respectively. The calculation were first performed  for equal intensity of $2\times{10}^{9}$  W/$\textrm{m}^{2}$ of the trap and the control lasers. To determine the wavelengths leading to state-independent trapping, the AC Stark shifts of the ground and the excited state of the $F_{g}=4\rightarrow F_{e}=5$ D$_2$ line transition were evaluated as functions of the trap laser wavelength for the two configurations of interest: $\lambda_{control}=2\lambda_{trap}$ and $\lambda_{\rm c}=3\lambda_{\rm t}$. Results of the calculations are shown in Figure \ref{fig lcont=2ltrapeqint1}. As expected, the excited state exhibits splittings whereas the ground state remains degenerate. From the numerical results of Fig. \ref{fig lcont=2ltrapeqint1}  we identify the following combinations of magic wavelengths of the trap and the control lasers: $931.8-1863.6$ nm for the case $\lambda_{\rm c}=2\lambda_{\rm t}$ and $927.5-2782.5$ nm for $\lambda_{\rm c}=3\lambda_{\rm t}$ respectively.

\begin{figure}
\begin{center}
\resizebox*{8.5cm}{!}{\includegraphics{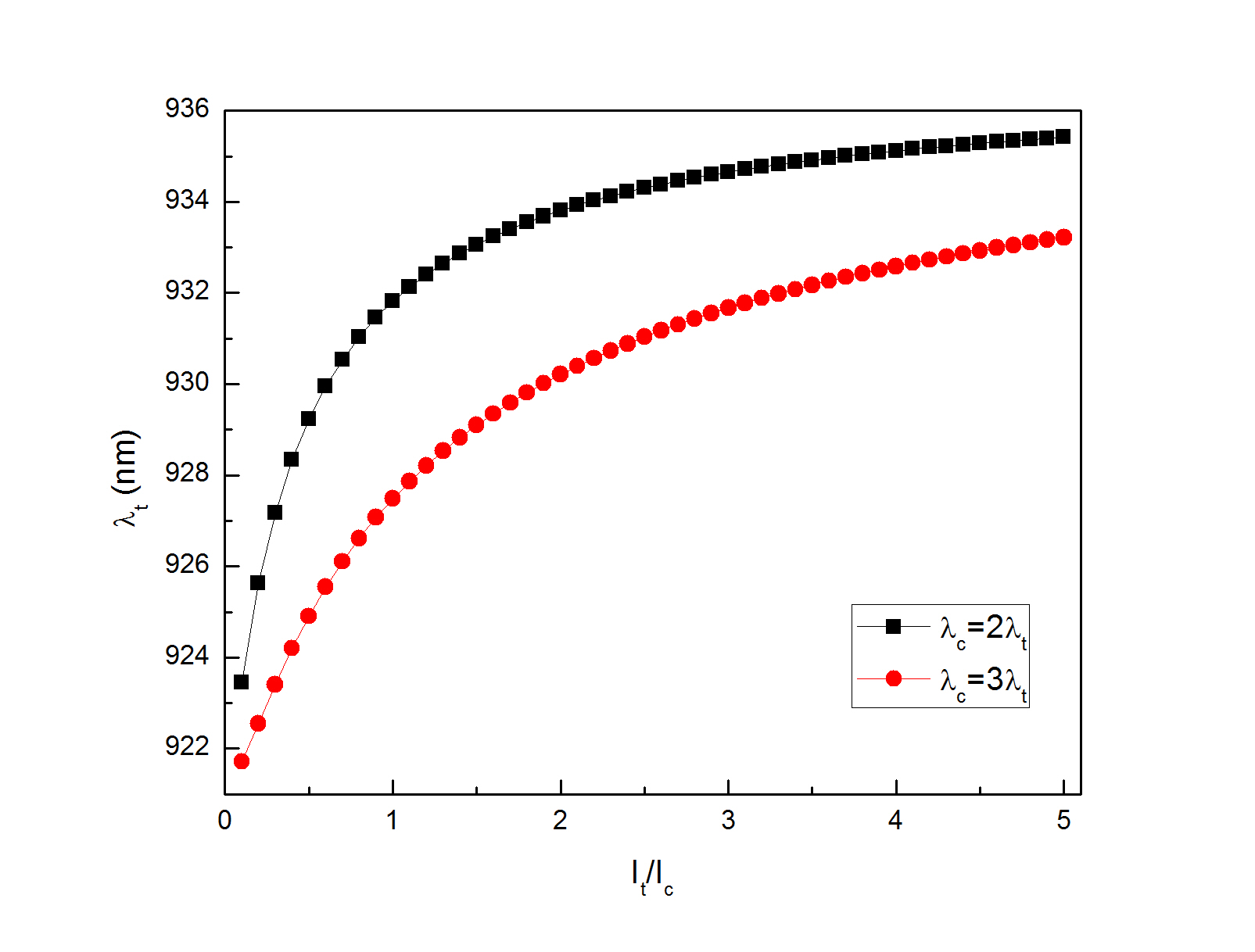}}
\caption{Trap laser wavelength required for state-independent trapping versus the ratio of trap and control laser intensities. The two data sets correspond to the cases $\lambda_{control}=2\lambda_{trap}$ and $\lambda_{control}=3\lambda_{trap}$.  The lines  are guides for the eyes.}
\label{fig_rap}
\end{center}
\end{figure}

Moreover, it is worth noting that, in the case of bichromatic trapping, the wavelength combination leading to state-independent trapping depends on the ratio of the trap and control laser intensities, thus introducing a degree of tunability of the magic wavelength.  Such an the effect is illustrated in Figure \ref{fig_rap}, where the results of the calculation of the magic wavelength as a function of the intensity ratio of the trap and control lasers are reported for both $\lambda_{\rm c}=2\lambda_{\rm t}$ and $\lambda_{\rm c}=3\lambda_{\rm t}$ . The data indeed demonstrate  tunability of the magic wavelength in excess of 10 nm, by changing the relative intensities of the lasers. Thus, bichromatic trapping significantly extends the range of possible  combinations of magic wavelengths with respect to the monochromatic case.  

%
%

\subsection{Magic Wavelength Pairs  corresponding to available High Power Lasers}

As previously pointed out, our main motivation for investigating bichromatic state-insensitive trapping of caesium atoms is the absence of a laser  source at the magic wavelength of $935.6$ nm with sufficiently high power for atom trapping in free space. We thus here investigate combinations of magic wavelength which can be produced by high power lasers which are currently available, in the sense of there being a suitable laser gain medium that provides high power at the required wavelength.
%
%
\begin{figure}[hb]
\begin{center}
\begin{minipage}{130mm}
\subfigure{
\hspace*{-1.75cm}\resizebox*{7.75cm}{!}{\includegraphics{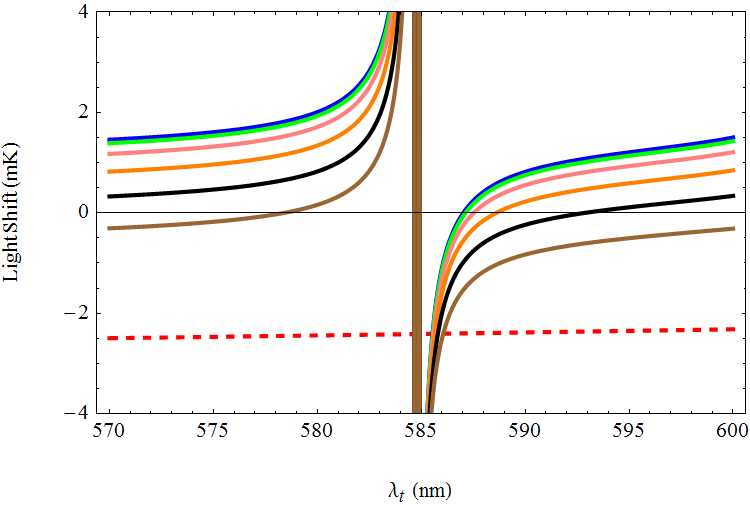}}}\label{fig p2}\hspace{6pt}
\subfigure{
\resizebox*{7.75cm}{!}{\includegraphics{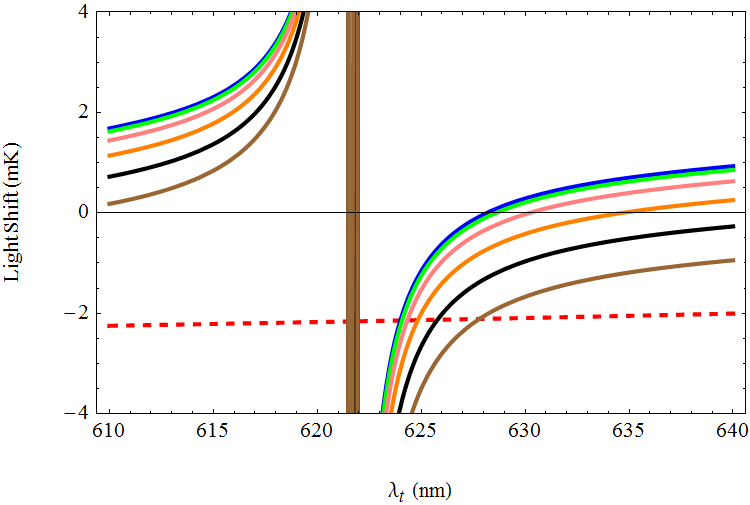}}}\label{fig p3}
\caption{AC Stark shift of ground and excited states of the $6S_{1/2}$ $F_{g}=4\rightarrow6P_{3/2} F_{e}=5$ transition as functions of the trap laser wavelength. The control laser wavelength is $1064$ nm and the power of both trap and control beams is $2$W. The red dashed line refers to the ground state $6S_{1/2}$ and the other bold lines refer to the Zeeman sublevels of the excited state $6P_{3/2}$, with the same color code as in Fig. 1. The left and right panel evidence two different regions of the examined wavelength range, with each of the regions containing a value of trap laser wavelength producing state-independent trapping.}
\label{fig day}
\end{minipage}
\end{center}
\end{figure}

For all  calculations presented in the following of the paper the beam waists of both trap and control lasers are equal to $10$ $\mu$m. The possible control laser wavelengths appropriate to achieve state-insensitive trapping were found to be $1064$ nm, $1070$ nm, $1075$ nm and $1080$ nm that can be achieved by Nd:YAG and ytterbium fiber lasers. The corresponding magic trap wavelengths accessible by high power lasers for all these control laser wavelengths were found in range of $585-588$ nm and $623-629$ nm which can be obtained by using orange and red Raman fiber lasers, in which a Yb-fiber laser generates stimulated Raman scattering in a second fiber, with the resulting infrared Raman emission frequency-doubled into the orange/red region of the visible.

The calculations of the AC Stark shift of the $6S_{1/2}$ $F_{g}=4\rightarrow6P_{3/2}$ $F_{e}=5$ transition for the control laser wavelength of $1064$ nm is presented in Figure \ref{fig day}. Two magic trap wavelengths were identified: $586.1$ nm and $627.8$ nm. 

\begin{figure}
\begin{center}
\begin{minipage}{130mm}
\hspace*{-1.5cm}\subfigure{
\resizebox*{8.5cm}{!}{\includegraphics{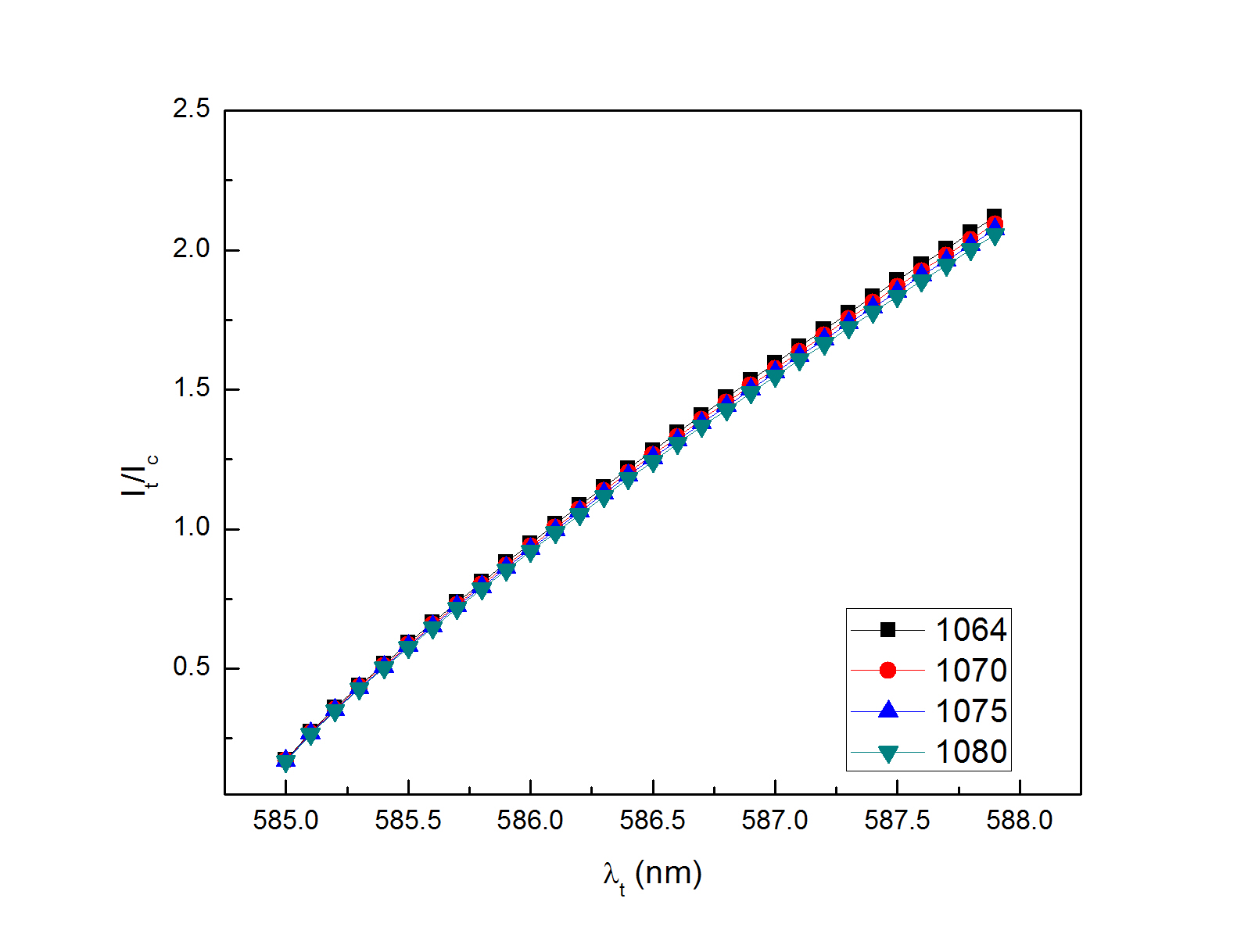}}}\hspace{6pt}
\subfigure{
\resizebox*{8.5cm}{!}{\includegraphics{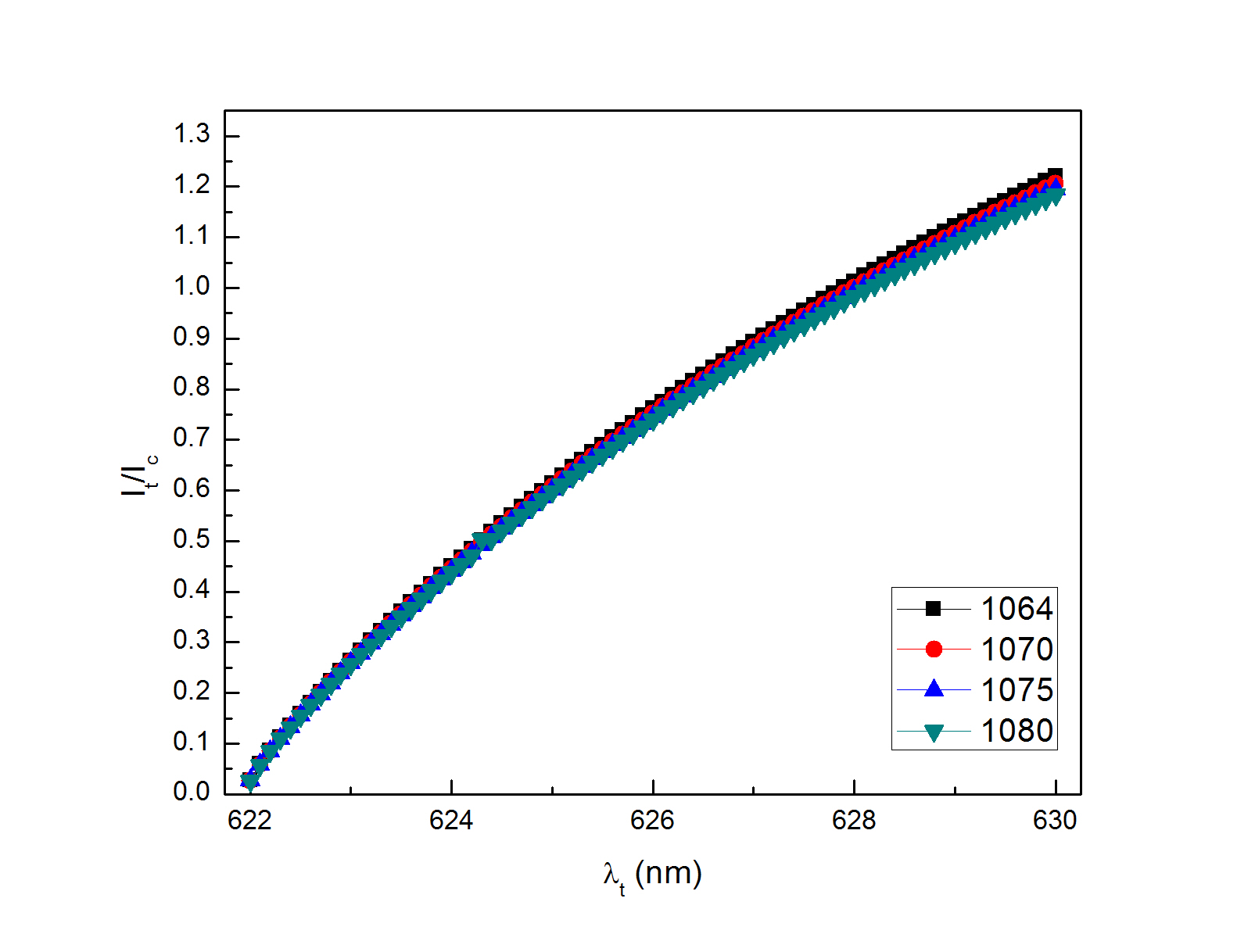}}}
\caption{Dependence of the ratio between the trap and the control lasers intensities on the wavelength of the trap laser, as required to maintain state-independent trapping. Different data sets correspond to different control laser wavelength of $1064$ nm, $1070$ nm, $1075$ nm and $1080$ nm. The left panel refers to  state-insensitive trapping obtained with a trap laser wavelength in the range of $585-587.9$ nm, the right panel refers to state-insensitive trapping obtained with a trap laser wavelength in the range of $623-630$ nm.}
\label{fig day1}
\end{minipage}
\end{center}
\end{figure}

\begin{figure}
\begin{center}
\begin{minipage}{130mm}
\hspace*{-1.5cm}\subfigure{
\resizebox*{8.5cm}{!}{\includegraphics{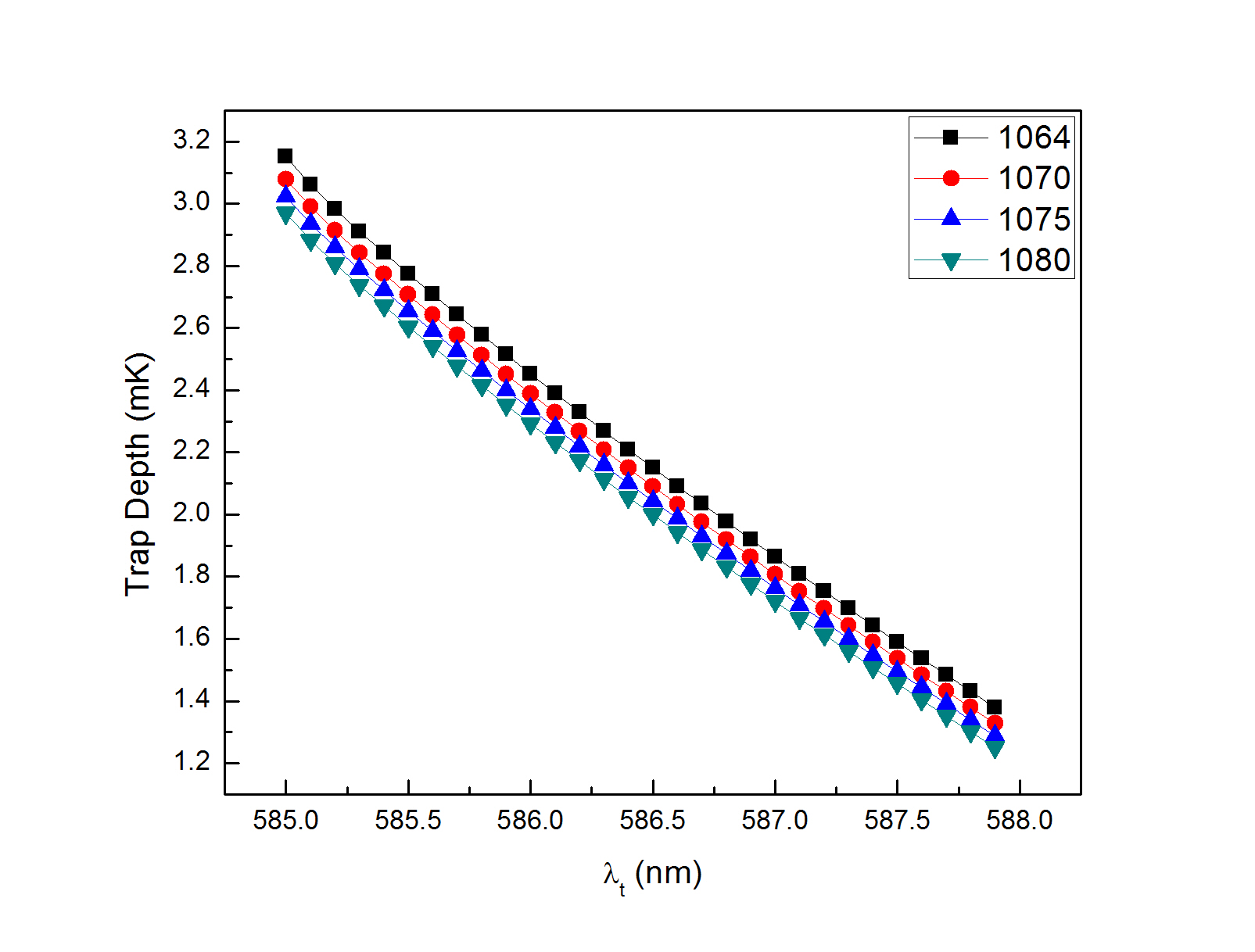}}}\hspace{6pt}
\subfigure{
\resizebox*{8.5cm}{!}{\includegraphics{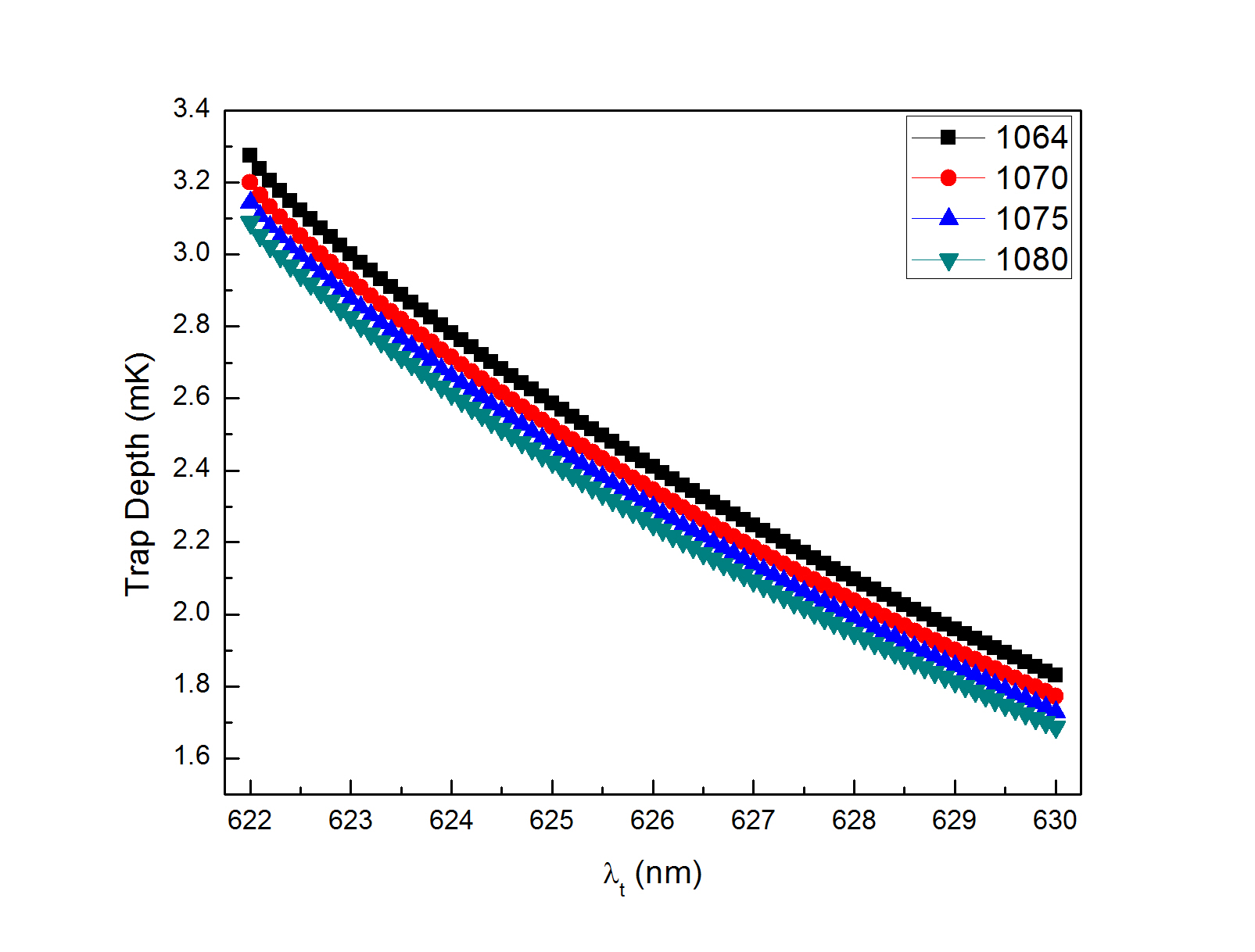}}}
\caption{Variation of obtained trap depth with trapping laser wavelength. All data correspond to state-independent trapping, as obtained by appropriately tuning the  ratio between the trap and the control lasers intensities.
Different data sets correspond to different control laser wavelength of $1064$ nm, $1070$ nm, $1075$ nm and $1080$ nm.}
\label{fig day3}
\end{minipage}
\end{center}
\end{figure}

The calculations presented in the previous Section  pointed out that, in the case of bichromatic trapping, it is possible to tune the  combination of magic wavelength of the trap and control lasers by varying their intensity ratio.  Such a tunability applies also to the combinations of laser wavelength considered in this Section.  The dependence of the required ratio between the trap and the control laser intensities on the wavelength of the trap laser is shown in Figure \ref{fig day1} for control laser wavelength of $1064$ nm, $1070$ nm, $1075$ nm and $1080$ nm.  For all of the four control laser wavelengths, the required intensity ratio of the trap and the control lasers exhibits an approximately linear dependence on the corresponding wavelength of the trap laser, thus offering precise tunability over a wide range of wavelengths. Nevertheless, there is a limit on the possible tuning range of  the magic wavelength. Figure \ref{fig day1} shows that  as the  trap laser wavelength increases, the intensity of the trap laser relative to the control laser also has to increase to maintain the condition of state-independent trapping. However, the increase in trap laser detuning decreases the trap depth. This is quantified in Figure  \ref{fig day3}, which evidences that the magnitude of the trap depth is inversely proportional to the trap wavelength. This sets a limit to the range of trap wavelengths usable for state-independent trapping. 
%
%
%
%

\begin{figure}
\begin{center}
{\resizebox*{8.5cm}{!}{\includegraphics{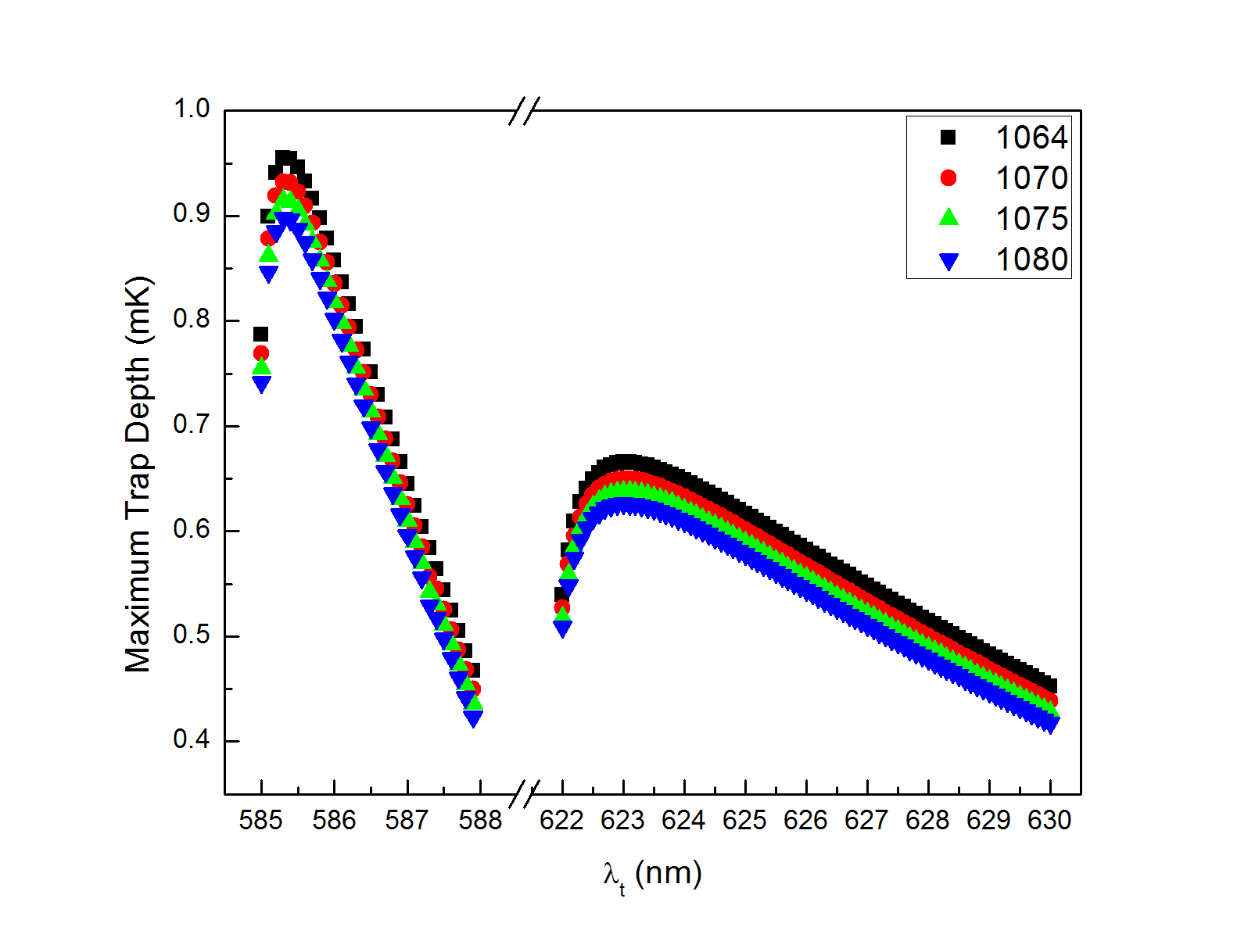}}}
\caption{Trap depth versus wavelength of the trap laser for a splitting between excited state sublevels equal to $2$$\Gamma$. The condition of state-independent trapping is satisfied for all reported data. The power of the control laser used in the calculation is $0.968$ W.}
\label{fig intradet}
\end{center}
\end{figure}

An important aspect for practical applications is the lifting of the excited state degeneracy produced by the trapping light.  As  Figure \ref{fig day} indicates, the excited state Zeeman sublevels are in general split, with the amount of splitting dependent on the trapping light wavelengths combination. Thus, each sublevel has a different combination of  trap/control magic wavelength. This poses limitations on the use of state-independent traps in conjunctions with mechanisms which may require efficient optical pumping. This is, for example, the case of optical molasses during the loading of a dipole trap from a magneto-optical trap. We thus investigated in detail the splitting between sublevels as a function of the experimental parameters, within the subset of these parameters which lead to state-independent trapping. We are in particular interested in determining the conditions for the splitting between the sublevels to be less than $-2$$\Gamma$, with $\Gamma$ the atomic excited state linewidth, which is a typical detuning for optical molasses. As the trap depth increases, in fact, the sublevels split further. Thus, to maintain the splitting between the sublevels below $2\Gamma$, the trap depth should be kept under a certain value. To determine such a value, the trap depth was evaluated as function of the magic trap wavelength for all control laser wavelengths for a fixed $2$$\Gamma$ splitting between the sublevels, with results in Figure \ref{fig intradet}. It follows that, with the proposed laser combinations, the deepest trap achievable to be able take advantage of simultaneous laser cooling and dipole trapping is $0.95$ mK for a trap laser wavelength of 585.3 nm.

\begin{figure}[h]
\begin{center}
\begin{minipage}{130mm}
\hspace*{-1.5cm}\subfigure{
\resizebox*{8.5cm}{!}{\includegraphics{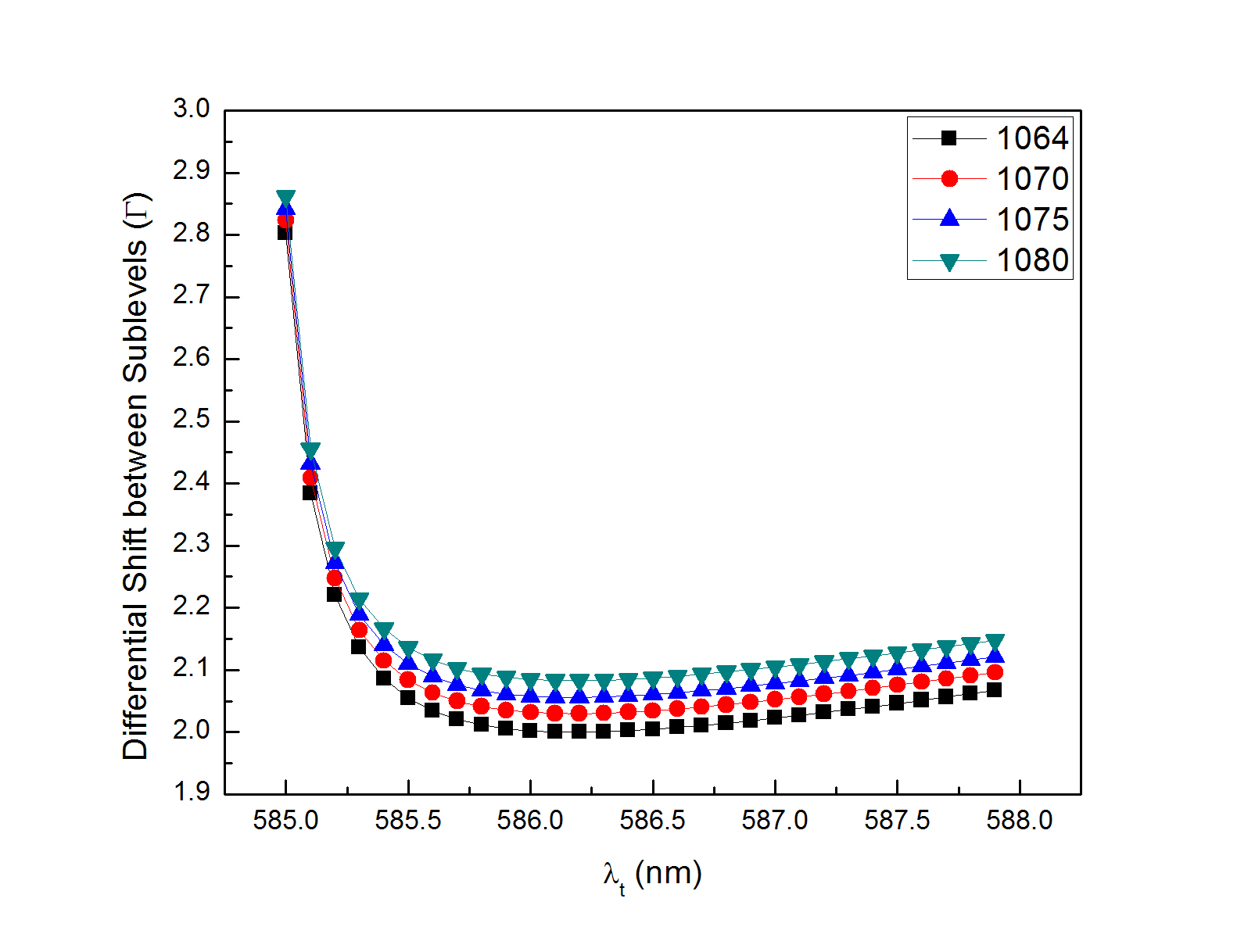}}}\hspace{6pt}
\hspace*{-0.5cm}\subfigure{
\resizebox*{8.5cm}{!}{\includegraphics{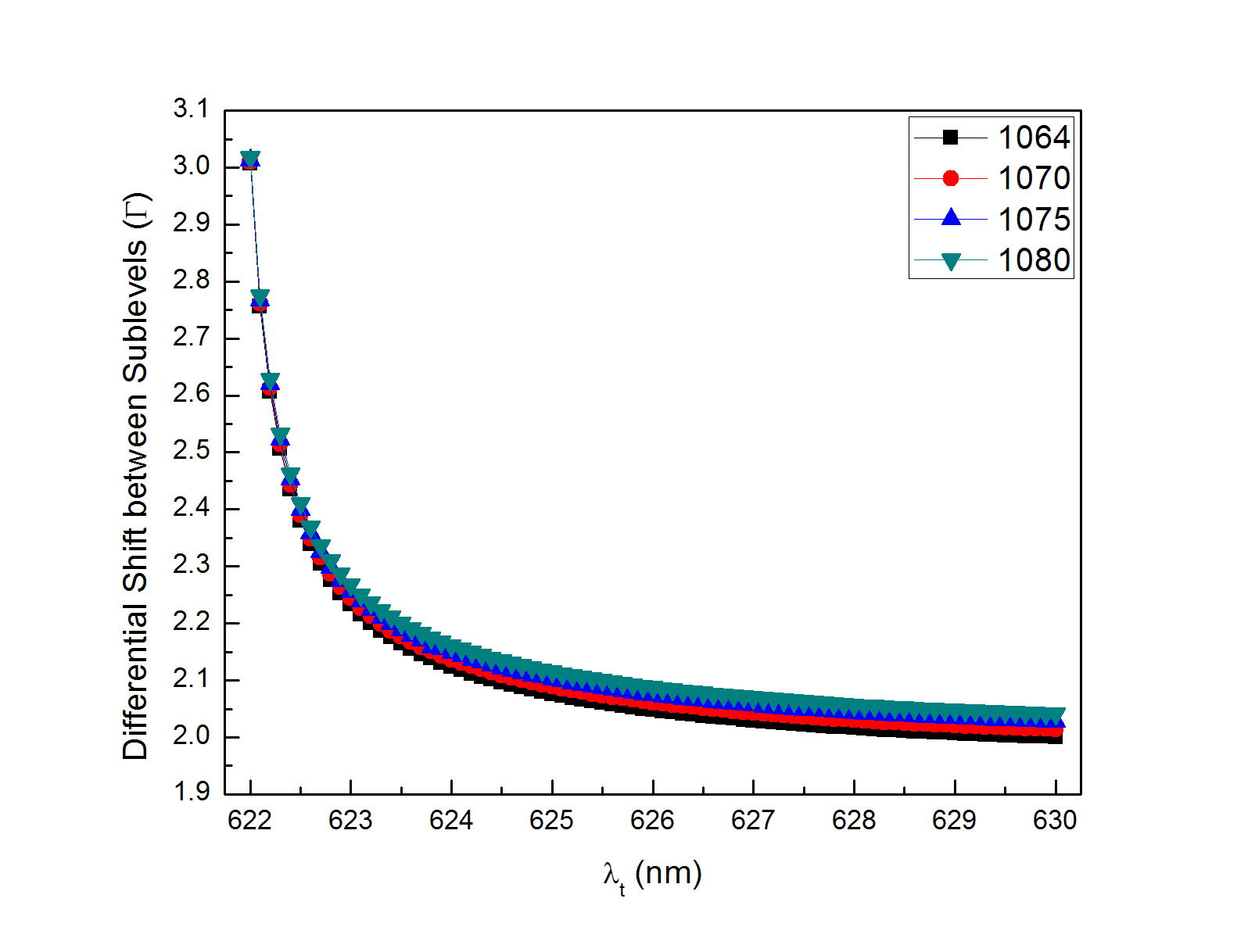}}}
\caption{Dependence on the trap laser wavelength of the splittings between the Zeeman sublevels of the $6P_{3/2}$ excites state for different control laser wavelengths of $1064$ nm, $1070$ nm, $1075$ nm and $1080$ nm.  The spilitting is reported in units of the excited state linewidth. The power of the control laser used in the calculation is  0.7W and 0.495W respectively for panel (a) and (b), respectively. }
\label{fig day5}
\end{minipage}
\end{center}
\end{figure}

In addition, the dependence on the trap laser wavelength of the splitting between the Zeeman sublevels was investigated, with results in Figure \ref{fig day5}. The actual value of the  combination of magic wavelength affects the magnitude of the splitting between the Zeeman sublevels. The lowest splitting occurs at magic wavelength pairs of $586.2-1064$ nm and $630-1064$ nm. As the data indicates, any deviation from these values leads to an increase in the magnitude of the splitting. Given the very different trap depth obtainable at these different wavelength, the magic wavelength pair  $586.2-1064$ nm is more favourable for state-independent trapping.

\subsection{Effects of  laser intensity and wavelength variations}

\begin{figure}[h]
\begin{center}
\begin{minipage}{130mm}
\centerline{\subfigure{
\resizebox*{8cm}{!}{\includegraphics{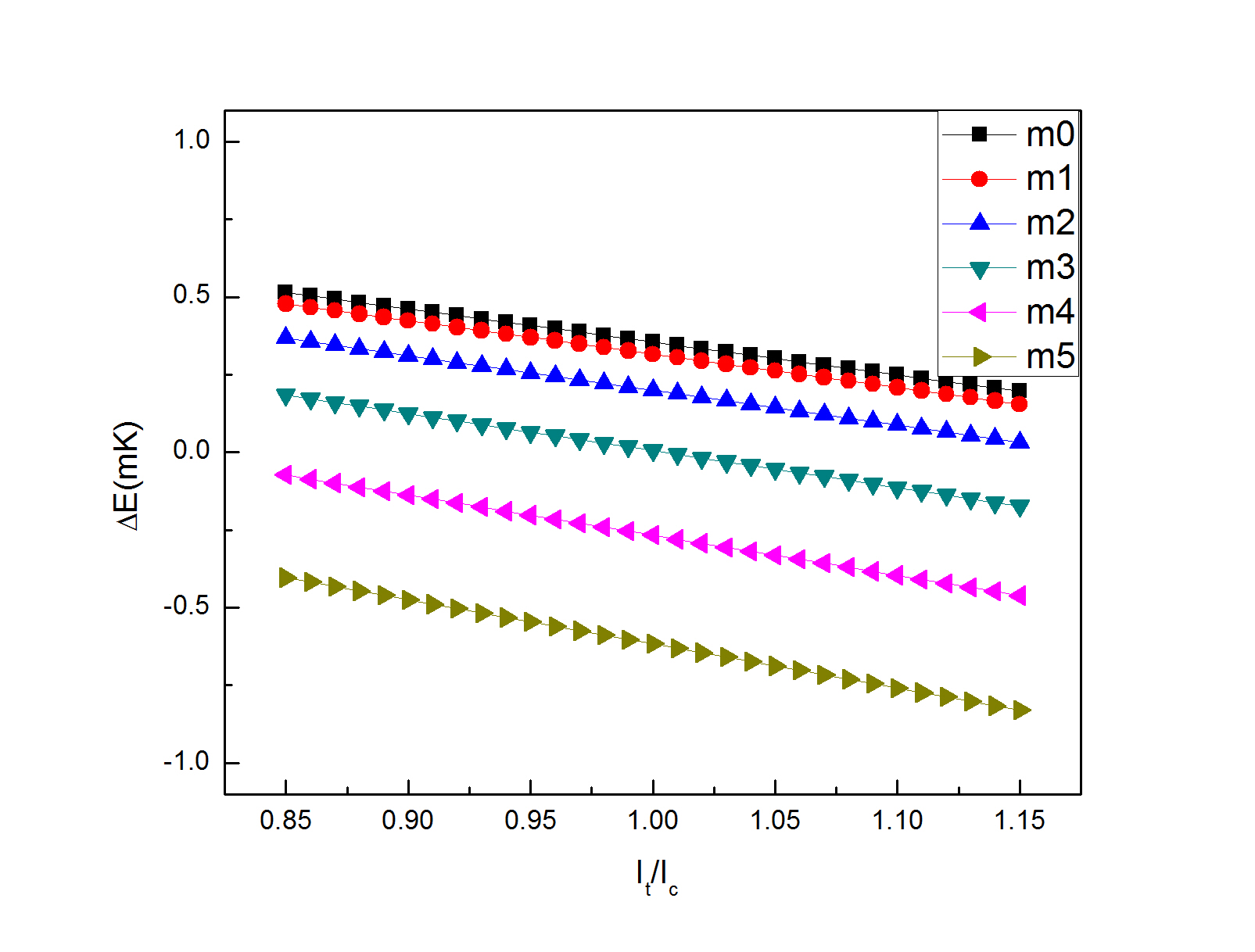}}}}
\caption{Differential AC Stark shift produced by laser intensity variations. Individual differential AC Stark shifts of the different excited state sublevels are reported.  The power of the control laser used in the calculation is $2$ W.}
\label{fig:intfluc}
\end{minipage}
\end{center}
\end{figure}

In a bichromatic state-insensitive trap,  variations in laser intensity or wavelength lead to a violation of the conditions for state-insentive trapping. The effect of these variations is quantitatively investigated in this Section. Specifically, we investigated the effect of intensity variations of the trap and control lasers on the AC Stark shift experienced by the ground and the excited states for the magic wavelength pair of $623.5-1064$ nm and a trap depth of $1$ mK. The differential AC Stark shift was evaluated for a variation of  $\pm$$15\%$ in the intensity ratio of the trap and the control lasers. Different excited state Zeeman sublevels experience different shifts induced by the variations in laser intensity.
Figure \ref{fig:intfluc} shows that the different shift of the different Zeeman sublevels of the excited state vary linearly with the intensity ratio. Also we notice that a variation in the intensity ratio of lasers causes a variation in the magic wavelength
 of each Zeeman sublevel.  As the intensity ratio deviates from the value corresponding to state-independent trapping for a specific sublevel of the excited state, it approaches the required value of intensity ratio for the next sublevel.


\begin{figure}[h]
\begin{center}
\resizebox*{8cm}{!}{\includegraphics{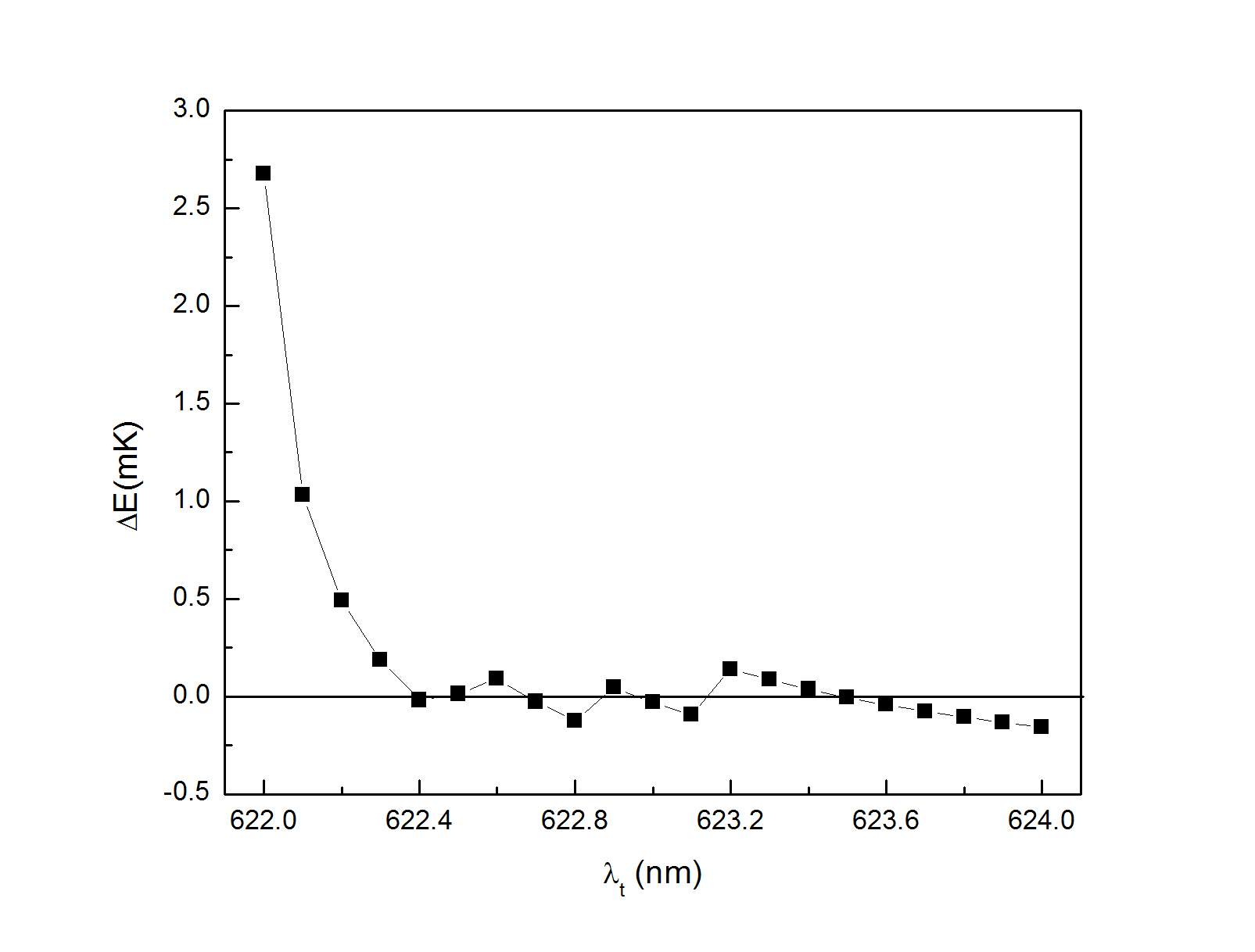}}
\caption{Differential AC Stark shift between excited and ground state of the caesium atom as a function of the trap laser wavelength. The differential light shift of the excited state sublevel which experiences the smallest shift is displayed. Calculations are for an intensity ratio $I_t/I_c = 0.36053$.  The power of the control laser used in the calculation is $2$ W.}
\label{fig lamfluc}
\end{center}
\end{figure}

The other experimental parameter whose variations affect the condition of state-independent trapping is the wavelength of the two lasers. This effect was investigated by considering the variations of the laser wavelengths from the required value. The differential AC Stark shift between the ground and the excited states of the caesium atom was calculated for a variation of the trap laser wavelength up to a value $\Delta\lambda_{t}=\pm1.5$ nm while keeping the intensity ratio fixed for a combination of magic wavelengths of $623.5-1064$ nm. The differential light shift corresponding to the excited state sublevel which experiences the smallest shift was calculated. Results of the calculations are reported in Figure \ref{fig lamfluc}, where the effects of the trap laser wavelength variations are quantified. We notice that as the trap laser wavelength approaches the $8D_{3/2}$ transition at $621.7$ nm the differential AC Stark shift increases significantly. Away from this transition, as the the trap laser wavelength increases the differential AC Stark shift exhibit an oscillatory value around zero.

\section{Conclusion}

In this work bichromatic state-insensitive trapping of caesium atoms was investigated to overcome the problem of lack of sufficiently intense laser sources for free space monochromatic trapping.  Our results identified several  combinations of magic wavelengths promising for practical realizations. First,  combinations of magic wavelengths were determined as $931.8-1863.6$ nm and $927.5-2782.5$ nm for the specific cases of $\lambda_{\rm c} = 2\lambda_{\rm t}$ and $\lambda_{\rm c} = 3\lambda_{\rm t}$ with equal power. The required laser beams may be derived from a single laser source via second- or third-harmonic generation. Then magic wavelength pairs achievable by two currently available high power lasers were identified. We found that for a control laser wavelength of $1064$ nm, $1070$ nm, $1075$ nm and $1080$ nm the corresponding magic trap wavelength accessible by high power lasers for all these control laser wavelengths were in the range of $585-588$ nm and $623-629$ nm. These wavelength combinations can be implemented by employing a red or orange Raman fiber laser as trap laser and a Nd:YAG or ytterbium fiber laser as control laser. 

This work demonstrates that bichromatic trapping extends the range of possible magic wavelengths compared to the specific magic wavelength associated to a single laser scheme. Bichromatic trapping also introduces a degree of tunability of the magic wavelength, as by varying the trap/control laser intensity ratio it is possible to vary the wavelength combination leading to state-independent trapping. Therefore the limitations imposed on the power and wavelength of the laser source in the case of the single laser scheme are overcome by bichromatic trapping. In this perspective, bichromatic schemes for state-insensitive trapping are more advantageous than single laser schemes.   

\section*{Acknowledgements}

We thank Dr Luca Marmugi (UCL) for critical reading the manuscript. This work was supported by EPSRC (Grant No. EP/H049231/1)

\end{document}